\documentclass[twocolumn]{article}

\pdfoutput=1

\usepackage{graphicx}
\usepackage{amsmath}
\usepackage{authblk}

\title{Monte Carlo simulations in anomalous radiative transfer: a tutorial}

\author[1,2,*]{Tiziano Binzoni}
\author[3]{Fabrizio Martelli}

\affil[1]{Department of Basic Neurosciences, University of Geneva, Geneva, Switzerland}
\affil[2]{Department of Radiology and Medical Informatics, University Hospital, Geneva, Switzerland}
\affil[3]{Dipartimento di Fisica e Astronomia dell'Universit\`a degli Studi di Firenze, Sesto Fiorentino, Firenze, Italy}

\affil[*]{Corresponding author: tiziano.binzoni@unige.ch}

\begin{document}

\maketitle

\begin{abstract}
Anomalous radiative transfer (ART) theory is a generalization of classical radiative transfer theory.
The present tutorial wants to show how Monte Carlo (MC) codes describing photons transport
in anomalous media can be implemented.
It is shown that the heart of the method consists in suitably describing, 
in a ``non-classical'' manner, photons steps starting from fixed light sources or from  
boundaries separating regions of the medium with different optical properties.
To give a better intuition of the importance of these particular photon step lengths,
it is also shown numerically that the described approach is essential to preserve 
the  invariance property for light propagation.
An interesting byproduct of the MC method for ART, 
is that it allow us to simplify the structure of ``classical'' MC codes,
utilized e.g. in biomedical optics.
\end{abstract}

\section{Introduction}
\label{sec:Introduction}

Monte Carlo (MC) simulations represent the tool of choice when describing
``classical'' photons migration in propagating media \cite{ref:ZhuLiu2013}.
This is specially true when exact solutions (except for statistical noise) of complex media are needed.
In fact, when dealing with the simultaneous presence of different media types, complex geometries 
and varying optical parameters, it becomes extremely difficult to use alternative approaches such us analytical solutions
or finite differences based methods.
Moreover, the continuous increase of the computational power, 
even for simple desktops, renders in general more and more attractive and easy to use the MC methods.
These are some of the reasons why data generated by MC simulations have become nowadays a ``gold standard'',
allowing to test alternative methods.

In recent years, a generalization of the classical theory  
allowing  also to describe anomalous radiative transfer (ART), has been introduced
\cite{ref:LarsenVasques2011,ref:rukolaine2016}. 
ART theory permits to modelize photons migration in anomalous media, i.e., media where the classical 
Beer–Lambert–Bouguer's law is not valid.

The release of the Beer–Lambert–Bouguer's law hardly complexify the definition of the analytical models
describing ART and their related solutions. 
This increase in complexity may be generated, e.g., by the need of the introduction of models based on  fractional (integro)-differential equations, 
where the solutions are often difficult to obtain and manipulate. 
Moreover, when the investigated optical medium is not infinite, and thus
necessitates the introduction of boundary conditions,
the definition of the analytical models and the obtainment of their solutions may represent a real challenge.

This is why in communities such as the biomedical optics, where the main aim is not specifically to understand the mathematical theory but above all to obtain solutions related to complex practical problems in photon migration, more efficient and general purpose tools would be particularly appreciated.
It is precisely in the context of  ART that the MC methods show their extreme power, i.e., 
thanks to their relative simplicity in describing and solving the considered problems.

In general, the MC methods describe how photons propagate step-by-step in the media,
through scattering, absorption, reflection and refraction events.
The main difference between the classical and the ART 
MC-approaches is given by the specific choice
of the function describing the probability $p(s)ds$ for a photon to reach a distance   
$s\in[s-ds,s+ds]$  in the medium, without interactions; where
$p(s)$ is a probability density function (pdf).

As it is well known, in the classical case
\begin{equation}
p(s)=\frac{1}{\ell}e^{-s/\ell},
\label{eq:classicalpdf}
\end{equation}
where $\ell$ is the mean free path for the investigated medium (in biomedical optics $\ell$ is often reported as
$1/\ell=\mu_t$; where $\mu_t$ is the extinction coefficient).
Historically, Eq. (\ref{eq:classicalpdf}) has been determined by considering the fact that a classical medium
must satisfy Beer–Lambert–Bouguer's law.

In ART,  it is not mandatory for $p(s)$ to be an exponential function
and a panoply of choices, depending on the specific optical properties
of the medium, are possible.

At this point, the fundamental remark is that to simulate ART models, by means of
the MC method, {\it it is not sufficient to substitute in the MC code the classical $p(s)$
[Eq. (\ref{eq:classicalpdf})] with the relative anomalous one}.
In fact, if we proceed in such a simple way the 
fundamental reciprocity law (RL) at the
basis of optics will not be correctly reproduced by the simulations.    
As a short reminder, in the present context the RL tells us that if we exchange light sources and detectors
(by inverting the direction of the detected photons; 
see e.g. red line in Fig. \ref{fig:figureTwoSperes}), we must always measure (detect) the same photons flux.
For a more indepth approach to the RL  see e.g. Ref. \cite{ref:CaseZweifel67}.

Thus, based on the findings derived from ART theory, the aim of the present tutorial
is to describe the suitable way to treat this problem.
This will be done by directly giving the necessary practical ``recipes'' allowing to build
MC codes for ART.  
We will see that the core of the solution is given by the particular choice of the pdf allowing  to generate the length $s$ of the {\it first step}
of photons that are {\it starting from a medium boundary} 
(defined by adjacent regions with different optical parameters) {\it or from a fixed light source}. 
The remaining steps, and the photons interactions with absorbers/scatterers, being treated as in a classical MC code.

In this tutorial, it is considered that the reader already has a basic knowledge on 
MC simulations applied to classical problems in photons migration, typical e.g. for biomedical optics.
Very simplified numerical  examples will be given, allowing to
intuitively understand the influence of a correct/incorrect approach. 

To simplify the explanations, even if ART theory clearly  include the classical case,
in the following sections {\it we will sometimes adopt the convention that the expression ART 
does not include the classical case}.
The context easily clarifies this use.

\section{Basic MC tools in ART}
\label{sec:Basic MC tools in ART}

\subsection{Basic probabilty density functions for photons steps}
\label{sec:Basic probability density functions}

ART theory is a {\it generalization} of the classical 
radiative transfer theory.
We will 
recall here how in ART photons steps are generated
and, in particular, how the interactions with the boundaries and light sources are treated \cite{ref:dEon2018,ref:dEon2018b}.

In ART, photons steps, $s_c$, occurring  inside the medium are generated by giving a pdf 
\begin{equation}
p_c(s_c;a,b,\dots),
\end{equation}
defining the probability $p_c(s_c;a,b,\dots)ds_c$ to reach a distance 
$s_c\in[s_c-\frac{ds_c}{2},s_c+\frac{ds_c}{2}]$ in the investigated anomalous medium.
Thus, the generalization of the classical radiative tranfer theory consists in the fact 
that $p_c(s_c;a,b,\dots)$ must not necessarily have an exponential behavior [Eq. (\ref{eq:classicalpdf})]. 
In other words, ART can treat photons propagation in  media that are not memory-less.
The parameters $a,b,\dots$ represent the constants defining a specific pdf,
depending on the optical characteristics of the medium. 
The index $c$ is to remember that the positions of the sites where the scattering or absorbing events 
occur in the medium are {\it statistically correlated}. 
In fact, these positions can statistically be described by the use of a probability distribution function 
$v_k(s_c;a,b,\dots)$ which, in principle, can directly be derived from $p_c(s_c;a,b,\dots)$ 
(see Sec. \ref{sec:Basic MC tools in ART}.\ref{sec:Extracting the physical properties from the PDF}).
The specific choice of the model for $p_c(s_c;a,b,\dots)$ comes from the 
actual optical properties of the physical system (medium) we want to describe.
Thus, $p_c(s_c;a,b,\dots)$ can in principle be assessed theoretically or experimentally.

Until this point, the ART approach appears to be similar to the classical, i.e.,
once the suitable $p_c(s_c;a,b,\dots)$  is defined, we can generate the photons step lengths,
and decide if the photons are scattered, absorbed, reflected or refracted at the different interaction sites.
However, the fundamental difference clearly appears when a photon hits a boundary. 
In this case, if the boundary is encountered when traveling along the step $s_c$,
then in ART {\it the photon simply  stops at the boundary}.
From this reached position a {\it new step} is immediately generated, 
without the need to terminate the previous path by applying the well known procedure 
usually utilized in the classical case (see e.g. Ref. \cite{ref:Wangetal95}).

When we generate a new step $s_u$ starting from a {\it fixed} boundary or 
a {\it fixed} light source position
(i.e. these positions do obviously not result from the probability law $v_k(s_c;a,b,\dots)$), 
the pdf can no more be $p_c(.)$. 
In this case, a different pdf must be considered, i.e. \cite{ref:dEon2018},
\begin{equation}
p_u(s_u;a,b,\dots)=\frac{1-\int_0^{s_u} p_c(s_c;a,b,\dots)ds_c}{\int_0^{+\infty} s_c p_c(s_c;a,b,\dots)ds_c},
\label{eq:pupdf}
\end{equation}
where the index $u$ is to remember that the photon starts from an {\it uncorrelated origin}
(i.e. the position is fixed).
Thus, $p_u(s_u;a,b,\dots)ds_c$ represents the probability to reach a distance 
$s_u\in[s_u-\frac{ds_u}{2},s_u+\frac{ds_u}{2}]$, 
if starting from a fixed position (boundary or light source).
All remaining steps are generated with the law $p_c(s_c;a,b,\dots)$.

As we mentioned in Sec. \ref{sec:Introduction}, Eq. (\ref{eq:pupdf}) derives from the fact that, similarly to the classical case, ART 
must also satisfy the fundamental RL. 
In other words, if we do not take into account Eq. (\ref{eq:pupdf}), 
e.g. by simply putting $p_u(.)=p_s(.)$, 
the RL is no more satisfied (except for the classical case, see below)
and photons propagation in the medium will be described in a wrong manner.
Note that, Eq.  (\ref{eq:pupdf}) permits the RL to be satisfied for any choice of optical parameters (scattering coefficients, phase functions, refractive indexes, etc.) or geometries.

The random steps, $s_c$ and $s_u$, based on laws $p_c(s_c;a,b,\dots)$ and $p_u(s_u;a,b,\dots)$, can numerically be generated as usual
by analytically solving 
\begin{equation}
\xi=\int_0^{s_i} p_i(s_i';a,b,\dots)ds_i'; \quad i\in\{c,u\},
\label{eq:pupdfrepresentation}
\end{equation}
as a function of $s_i$, where $\xi\in(0,1)$ is a uniformly distributed random variable.
If an analytical solution of Eq. (\ref{eq:pupdfrepresentation})  does not exist, a numerical solution can be  found, and 
a look-up table relating $\xi$ to $s_i$ can be derived. 

This is all we need to know to generate an ART MC simulation,
the remaining part of the MC code
being the same as in the classical case.


\subsection{Extracting general medium properties from $p_c(s_c;a,b,\dots)$}
\label{sec:Extracting the physical properties from the PDF}

Once $p_c(s_c;a,b,\dots)$, the geometry, the optical parameters and the phase functions
are given, it becomes possible to perform MC simulations in the ART domain.
However, before to run a specific simulation, it is maybe interesting to know
that from  $p_c(s_c;a,b,\dots)$ we can derive some very general properties
of the investigated anomalous medium.
Strictly speaking, the properties presented here hold for an infinite medium
with pdf $p_c(s_c;a,b,\dots)$, but are of fundamental importance in understanding
the physics described by $p_c(s_c;a,b,\dots)$. 
Thus, for the sake of completeness, in the following paragraphs we will report four functions, directly derived from 
$p_c(s_c;a,b,\dots)$, describing some of these properties. 
The reported equations are derived from well known results in renewal theory and the interested  
reader can refer e.g. to Refs. \cite{ref:Mainardi2007,ref:Ross2019}.
The equations hold for a photon propagation through any homogeneous
part of the medium. 

1) The first interesting function is the survival probability $P_c(s_c;a,b,\dots)$, i.e., in the language of optics, 
the probability that a photon survive after a path of length $s_c$. 
This probability is expressed as
\begin{equation}
P_c(s_c;a,b,\dots) = \int_{s_c}^{+\infty }p_c(s_c';a,b,\dots) ds_c'.
\label{eq:Pc}
\end{equation}
In the classical case, $P_c(s_c;a,b,\dots)$ corresponds to the Beer–Lambert–Bouguer's law
(within a multiplicative factor).

2) Now, let be 
\begin{equation}
\tilde p_c(w_c;a,b,\dots)=\mathcal{L}\{p_c(s_c;a,b,\dots);w_c\},
\label{eq:pcL}
\end{equation}
where $\mathcal{L}\{.;.\}$ represents the Laplace transform and  $w_c$  the 
obtained variable in the transformed domain.
Then, we can express the average number of photon interactions along a path of length $s_c$ as 
\begin{equation}
\langle m(s_c;a,b,\dots) \rangle = \mathcal{L}^{-1}\left\{
\frac{\tilde p_c(w_c;a,b,\dots)}{w_c[1-\tilde p_c(w_c;a,b,\dots)]};s_c\right\};
\label{eq:m}
\end{equation}

3) The probability that $k$ photon interactions occur along a path of length $s_c$ is
\begin{multline}
v_k(s_c;a,b,\dots)=\mathcal{L}^{-1}\left\{\frac{1-\tilde p_c(w_c;a,b,\dots)}{w_c} \right. \\
 \left. \times [\tilde p_c(w_c;a,b,\dots)]^k;s_c\right\};
\label{eq:vk}
\end{multline}

4) The probability that the sum of the first $k$ photon steps does not exceed $s_c$ is 
\begin{equation}
F_k(s_c;a,b,\dots)=\mathcal{L}^{-1}\left\{\frac{[\tilde p_c(w_c;a,b,\dots)]^k}{w_c};s_c\right\}.
\label{eq:Fk}
\end{equation}
The above equations allow us to better describe the physical system under study,
determined by a given choice of $p_c(s_c;a,b,\dots)$.

\section{ART MC simulations: explanatory examples}
\label{sec:Anomalous transport: an explanatory example}

The scope of the following examples is to show the importance of Eq. (\ref{eq:pupdf}) 
(the core of the ART MC simulations).
This will be done by demonstrating, with easy to understand numerical examples, that
if we neglect Eq. (\ref{eq:pupdf}), a fundamental physical property 
of the system cannot be reproduced, i.e., the invariance property (IP) will not be satisfied (see  below).   
Note that, the IP represents a very powerful test 
allowing us to check the reliability of any MC code \cite{ref:Martellietal21b}. 
 
\subsection{Two general analytical models}
\label{sec:Analytical model}

We will define here two possible models  for  $p_c(s_c;a,b,\dots)$, i.e., the power law and the constant step models.

\subsubsection{Power law}

The first model we will consider is called a ``power law'' and is expressed as  
\begin{equation}
p_c(s_c;\ell,a)=\frac{a(a+1)\ell(a\ell)^a}{(a\ell+s_c)^{a+2}}; \quad a>0,
\label{eq:ExpModelc}
\end{equation}
where $\int_0^\infty p_c(s_c;\ell,a) ds_c=1$ and $\int_0^\infty s_c p_c(s_c;\ell,a) ds_c=\ell$.
The asymptotic limit ($s_c \rightarrow +\infty$) of Eq. (\ref{eq:ExpModelc}) is
\begin{equation}
p_c(s_c;\ell,a)\sim\frac{a(a+1)\ell(a\ell)^a}{s_c^{a+2}}; \quad s_c\gg 1.
\label{eq:asympt}
\end{equation}
The specific $s_c$ dependence  of Eq. (\ref{eq:asympt}) is particularly interesting because it implies that it may exist an
asymptotic diffusion limit of this anomalous model \cite{ref:FrankSun2018}
(related to a ``classical'' or a ``fractional'' diffusion equation, depending on the choice of $a$; see below).
Thus, this simple model [Eq. (\ref{eq:ExpModelc})] represents a kind unified summary 
for the classical and anomalous approach.
 
Then, the pdf $p_u(s_u,\ell,a)$ can be derived by applying Eq. (\ref{eq:pupdf}), to obtain
\begin{equation}
p_u(s_u,\ell,a)=\frac{1}{\ell}\left(\frac{a\ell}{a\ell+s_u}\right)^{a+1}.
\label{eq:ExpModelu}
\end{equation}
Finally, the algorithms allowing to generate random $s_c$ and $s_u$ values in the MC codes
can be obtained by introducing Eqs. (\ref{eq:ExpModelc}) and (\ref{eq:ExpModelu}) 
into Eq. (\ref{eq:pupdfrepresentation}), and by solving as a function of $s_c$ and $s_u$, respectively.
This gives
\begin{equation}
s_c = a\ell\left[(1-\xi)^{-\frac{1}{a+1}}-1\right],
\label{eq:steps}
\end{equation}
and 
\begin{equation}
s_u = a\ell\left[(1-\xi)^{-\frac{1}{a}}-1\right].
\label{eq:stepu}
\end{equation}

\subsubsection{Constant step}

The second model we will consider in this tutorial is called 
``constant step'', because it describes photons that ``jump'' always
with  steps of the same length $\ell$ , i.e.,
\begin{equation}
p_c(s_c;\ell)=\delta(s_c-\ell),
\label{eq:step}
\end{equation}
where $\int_0^\infty p_c(s_c;\ell) ds_c=1$ and $\int_0^\infty s_c p_c(s_c;\ell) ds_c=\ell$.

Equation (\ref{eq:pupdf}) is then written as
\begin{equation}
p_u(s_u;\ell)=\frac{1-\Theta(s_u-\ell)}{\ell},
\end{equation}
where $\Theta(.)$ is the Heaviside function.

The generating functions for $s_c$ and $s_u$ are then expressed as [Eq. (\ref{eq:pupdfrepresentation})]
\begin{equation}
\label{eq:conSt1}
s_c=\ell; \quad \forall \xi,
\end{equation}
\begin{equation}
\label{eq:conSt2}
s_u=\ell \xi.
\end{equation}

\subsection{Numerical MC examples: IP test}

Let's now define four different examples of MC simulations based on Eqs. (\ref{eq:ExpModelc}) and (\ref{eq:step}). In the following sub-sections we will give the geometry of the problem, all the necessary optical quantities, together with the light sources and detectors
positions.
The general properties (geometry independent) of the four chosen tutorial examples are treated, and a short summary concerning the IP is also reported.

\subsubsection{Monte Carlo simulations}
\label{sec:Monte Carlo simulations}

\begin{figure}[htbp]
\centering
{\includegraphics[width=0.7\linewidth]{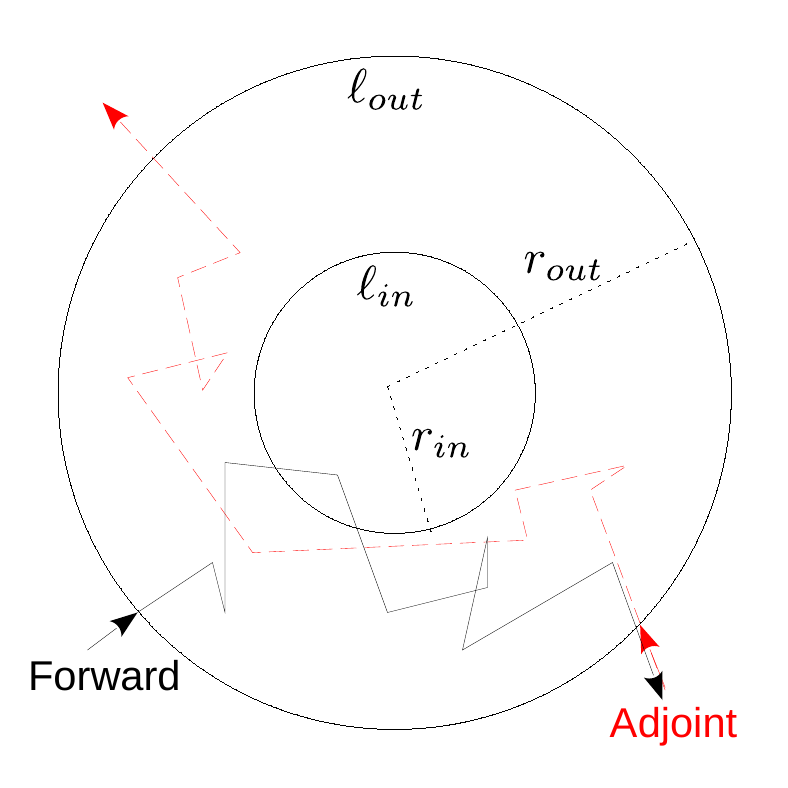}}
\caption{Schematic of the geometrical model utilized for the Monte Carlo simulations, representing two 3D spheres 
centered at the axis origin.}
\label{fig:figureTwoSperes}
\end{figure}

{\bf Geometry: }
The common geometrical model chosen in this tutorial consists
of two spheres centered at the axis origin, with radious $r_{in}<r_{out}$
and matched refractive indexes at all the boundaries
(Fig. \ref{fig:figureTwoSperes}). 
The mean free paths inside the spheres are $l_{in}$ and $l_{out}$. 
The absorption coefficients and the anisotropy parameters of the spheres are set to zero 
(mandatory conditions to test the IP in ART; see Sec. 
\ref{sec:Anomalous transport: an explanatory example}.\ref{sec:Invariance property}).

{\bf Forward measurements: } An isotropic and uniform radiance illumination impinging on the surface of the external sphere 
(continuous uniform distribution of Lambertian sources)
is utilized as a light source. 
Note that, in Fig. \ref{fig:figureTwoSperes} only one representative photon is reported (black path; forward propagation).
Photons are detected on the whole surface of the external sphere.
The pathlength of each single photon, $s$, is computed and the mean pathlength 
$\langle s \rangle^\textrm{Fwd}_\textrm{MC}$ is assessed.
Each simulation consists of 100 independent simulations (each of them has $10^7$ photons trajectories)
in order to evaluate the standard error of the mean.
Isotropic scattering is used.
Random photons steps for the three examples of power law are generated using Eqs. (\ref{eq:steps}) and (\ref{eq:stepu}), 
for three representative $a$ values, i.e., $a\in\{1,1/2,+\infty\}$ 
with
$\ell\in\{\ell_{in},\ell_{out}\}$ (the choice of $\ell_{in}$ or $\ell_{out}$
depends on where the photon is situated along the path).
For the constant step example, photons steps are obtained using 
Eqs. (\ref{eq:conSt1}) and (\ref{eq:conSt2}) with
$\ell\in\{\ell_{in},\ell_{out}\}$.

{\bf Adjoint measurements: } Adjoint measurements are assessed by inverting the direction of 
each photon reaching the sphere surface (detector).
Then, these photons, with their new inverted directions (red-dashed path in Fig. \ref{fig:figureTwoSperes}), 
are utilized as light source to generate the 
adjoint measurements, with the same conditions utilized for the forward measurements.
Even in this case, photons are detected on the whole surface of the external sphere.
At the end of the simulation, the mean pathlength, $\langle s \rangle^\textrm{Adj}_\textrm{MC}$,
is then computed.

{\bf The importance of $p_u(s_u,\ell,a)$: }
Forward and adjoint MC simulations are repeated two times: 
1) By using the rules for the MC ART presented in Sec. \ref{sec:Basic MC tools in ART}.\ref{sec:Basic probability density functions},
giving $\langle s \rangle^\textrm{Fwd}_\textrm{MC}$ and $\langle s \rangle^\textrm{Adj}_\textrm{MC}$ and;
2) By setting
$p_u(.)=p_c(.)$, as it is usually done in classical MC,
resulting in this case in two mean paths that we will note
$\langle s \rangle^\textrm{Fwd}_{\textrm{MC}_\textrm{no-reciprocity}}$ and 
$\langle s \rangle^\textrm{Adj}_{\textrm{MC}_\textrm{no-reciprocity}}$.

\subsubsection{Power law: $a=1$}
\label{sec:a=1}

In this case random steps $s_c$ and $s_u$ are generated as [Eqs. (\ref{eq:steps}) and (\ref{eq:stepu})]
\begin{equation}
s_c = \ell\left[(1-\xi)^{-\frac{1}{2}}-1\right],
\end{equation}
\begin{equation}
s_u = \frac{\ell\xi}{1-\xi}.
\end{equation}
To better understand the physical properties of this model, we will compare it with the classical one (see below),
through the functions presented in Sec.~\ref{sec:Basic MC tools in ART}.\ref{sec:Extracting the physical properties from the PDF}. 
In this case, Eq. (\ref{eq:Pc}) is expressed as
\begin{equation}
P_c(s_c;\ell,a=1) = \left(\frac{\ell}{\ell+s_c}\right)^2.
\label{eq:PPPP1}
\end{equation}
To obtain the remaining Eqs. (\ref{eq:m}), (\ref{eq:vk}) and (\ref{eq:Fk}) let first compute Eq. (\ref{eq:pcL}), i.e.,
\begin{equation}
\tilde p_c(w_c;\ell,a=1)=2 (w_c \ell)^2 \Gamma(-2,w_c \ell)e^{w_c \ell}.
\label{eq:pcLa1}
\end{equation}
where $\Gamma(.,.)$ is the incomplete gamma function.
Then, insert  Eq. (\ref{eq:pcLa1}) in Eqs. (\ref{eq:m}), (\ref{eq:vk}) and (\ref{eq:Fk}). 
Being the obtained expressions too complex to be derived analytically, 
numerical solutions can be assessed by applying the Talbolt's inverse Laplace transform \cite{ref:Abate2006} (see Fig. \ref{fig:FigRenewal}).

Note that, the asymptotic limit [Eq. (\ref{eq:asympt})] of this power law with $a=1$ can be described by a 
classical diffusion equation \cite{ref:FrankSun2018}. 

\subsubsection{Power law: $a=1/2$}
\label{sec:a=1/2}

The random steps $s_c$ and $s_u$ for $a=1/2$ are generated as  [Eqs. (\ref{eq:steps}) and (\ref{eq:stepu})]
\begin{equation}
s_c = \frac{\ell}{2}\left[(1-\xi)^{-\frac{2}{3}}-1\right],
\end{equation}
\begin{equation}
s_u = \frac{\ell}{2}\left[(1-\xi)^{-2}-1\right].
\end{equation}
This is an interesting case, because the asymptotic  limit of this model is related to a 
fractional diffusion equation with derivative of order $3/2$ \cite{ref:FrankSun2018}.
In this case [Eq. (\ref{eq:Pc})],
\begin{equation}
P_c(s_c;\ell,a=1/2) = \left(\frac{\ell}{\ell+2s_c}\right)^{\frac{3}{2}}
\label{eq:PPPP2}
\end{equation}
and Eq. (\ref{eq:pcL}) can be expressed as
\begin{equation}
\tilde p_c(w_c;\ell,a=1/2)=\frac{\sqrt{2}}{2}(\ell w_c)^{\frac{3}{2}}\Gamma(\frac{1}{2},\frac{\ell w_c}{2})e^{\frac{\ell w_c}{2}}
-\ell w_c +1
\end{equation}
Then, Eqs. (\ref{eq:m}), (\ref{eq:vk}) and (\ref{eq:Fk}) can be obtained numerically
with the same procedure presented in  
Sec. \ref{sec:Anomalous transport: an explanatory example}.\ref{sec:a=1}
 (see Fig. \ref{fig:FigRenewal}).

\subsubsection{Power law: $a=+\infty$}
\label{sec:a=Inf}

The random step $s_c$ and $s_u$ are generated as
\begin{equation}
s_c = s_u = -\ell\ln(1-\xi).
\end{equation}
In this case, $s_c = s_u=s$,
because the correlated and uncorrelated pdfs are the same, i.e.,
\begin{equation}
p_c(s;\ell,a)=p_u(s;\ell,a)=\frac{1}{\ell} e^{-s/\ell}.
\label{eq:classipdfsu}
\end{equation}
We immediately recognize here the pdf of the classical model,
related to the  underlying Beer–Lambert–Bouguer's law.
In fact [Eq. (\ref{eq:Pc})],
\begin{equation}
P_c(s_c;\ell,a=+\infty) = e^{-s_c/\ell}
\label{eq:PPPP4}
\end{equation}
It is in this sense, i.e., when $a=+\infty$,  that the classical model is naturally included in the  tutorial ART power model of
Sec.  \ref{sec:Anomalous transport: an explanatory example}.\ref{sec:Analytical model}.

In this special case,  by deriving before Eq. (\ref{eq:pcL}), i.e., 
\begin{equation}
\tilde p_c(w_c;\ell,a=+\infty)=\frac{1}{\ell w_c+1},
\end{equation}
Eqs.  (\ref{eq:m}), (\ref{eq:vk}) and (\ref{eq:Fk}) can be completely obtained analytically, i.e.,
\begin{equation}
\langle m(s_c;\ell,a=+\infty) \rangle =s_c/\ell
\label{eq:mcaInf}
\end{equation}
\begin{equation}
v_k(s_c;\ell,a=+\infty)=\frac{(s_c/\ell)^k}{k!}e^{-s_c/\ell}
\label{eq:vkaInf}
\end{equation}
\begin{equation}
F_k(s_c;\ell,a=+\infty)=\sum_{n=k}^{+\infty}\frac{(s_c/\ell)^n}{n!}e^{-s_c/\ell}
\end{equation}

\subsubsection{Constant step}
\label{sec:ConstantSteps}

By following the same procedure utilized for the power law, we obtain for the constant step model
\begin{equation}
P_c(s_c;\ell) = 1- \Theta(s_c-\ell),
\label{eq:PPPP3}
\end{equation}
\begin{equation}
\tilde p_c(w_c;\ell)=e^{-w_c \ell},
\end{equation}
\begin{equation}
\langle m(s_c;\ell) \rangle = \left \lfloor \frac{s_c}{\ell} \right \rfloor,
\end{equation}
where $\left \lfloor . \right \rfloor$ is the floor function and,
\begin{equation}
v_k(s_c;\ell)= \Theta(s_c- k \ell) - \Theta(s_c - (k+1)\ell),
\end{equation}
\begin{equation}
F_k(s_c;\ell)=\Theta(s_c- k \ell).
\end{equation}

\subsubsection{Invariance property (IP)} 
\label{sec:Invariance property}

As it was mentioned in 
Sec. \ref{sec:Basic MC tools in ART}.\ref{sec:Basic probability density functions},
the introduction of Eq. (\ref{eq:pupdf}) allows us to suitably taking into account the RL
in the MC ART simulations.  
Neglecting this rule, the RL may not be satified.
In this context, the examples chosen for this tutorial represent a special instructive case.
In fact, it is easy to see that, due to the absence of absorption and the particular geometry
of the light source and detector, the RL will {\it always} be satisfied, independently of the remaining optical parameters or other functions utilized for the MC simulation.
The reason of this special behavior is due to the fact that all the lunched photons 
always reach the detector (i.e., same photon flux through the detector in forward or adjoint direction).   
This means that testing the RL is not always a good test to check a MC code, 
and that a more universal method is needed.
This is why the IP is introduced in the following paragraph and is then exploited to show
the importance of the right choice of $p_u(.)$ [Eq. (\ref{eq:pupdf})].

The IP \cite{ref:BardsleyDubi1981,ref:Mazzolo2014} is a very powerful property that in the case of anomalous transport states the  following: let be 
a uniform and isotropic radiance illumination  
applied at the external boundary, $S$, of a finite, scattering and non-absorbing medium of any shape, volume V, matched refractive indexes and isotropic phase function.
Then, if we apply the rules reported in 
Sec. \ref{sec:Basic MC tools in ART}.\ref{sec:Basic probability density functions}
(to suitably describe ART), 
the mean pathlength, $\langle s \rangle_\textrm{theory}$, 
spent inside V by photons outgoing from $S$, is an {\it invariant quantity}
independent of the scattering strength.
Note that the volume $V$ can also be the union of any number of sub-volumes with different 
scattering coefficients but matched refractive indexes (external medium also matched).
The quantity $\langle s \rangle_\textrm{theory}$ is expressed as
\begin{equation}
\langle s \rangle_\textrm{theory}=4\frac{V}{S}.
\label{eq:IP}
\end{equation}
This property holds for anomalous and classical photon propagation and 
can be utilized to test the validity of the MC codes.
It is worth to note that in the case of classical photon transport
 (Beer–Lambert–Bouguer's law), the hypothesis 
of an isotropic phase function is not necessary.\cite{ref:BardsleyDubi1981,ref:Mazzolo2014,ref:Martellietal21}
A generalization of Eq. (\ref{eq:IP}) for unmatched refractive indexes also exists \cite{ref:Tommasi2020,ref:Martellietal21}, but this goes far from the aim of the 
present tutorial.
However, in this context we need a general formulation valid for ART,
implying the use of an isotropic phase function.

\section{Results} 
\label{sec:Results}

\subsection{General comparisons of the classical and the ART models} 

Before to proceed with the MC simulations, let us compare (semi)-analytically 
the power law and the constant step models with the classical one
[Eq. (\ref{eq:classipdfsu})].
 
\begin{figure}[t]
\centering 
{\includegraphics[width=\linewidth]{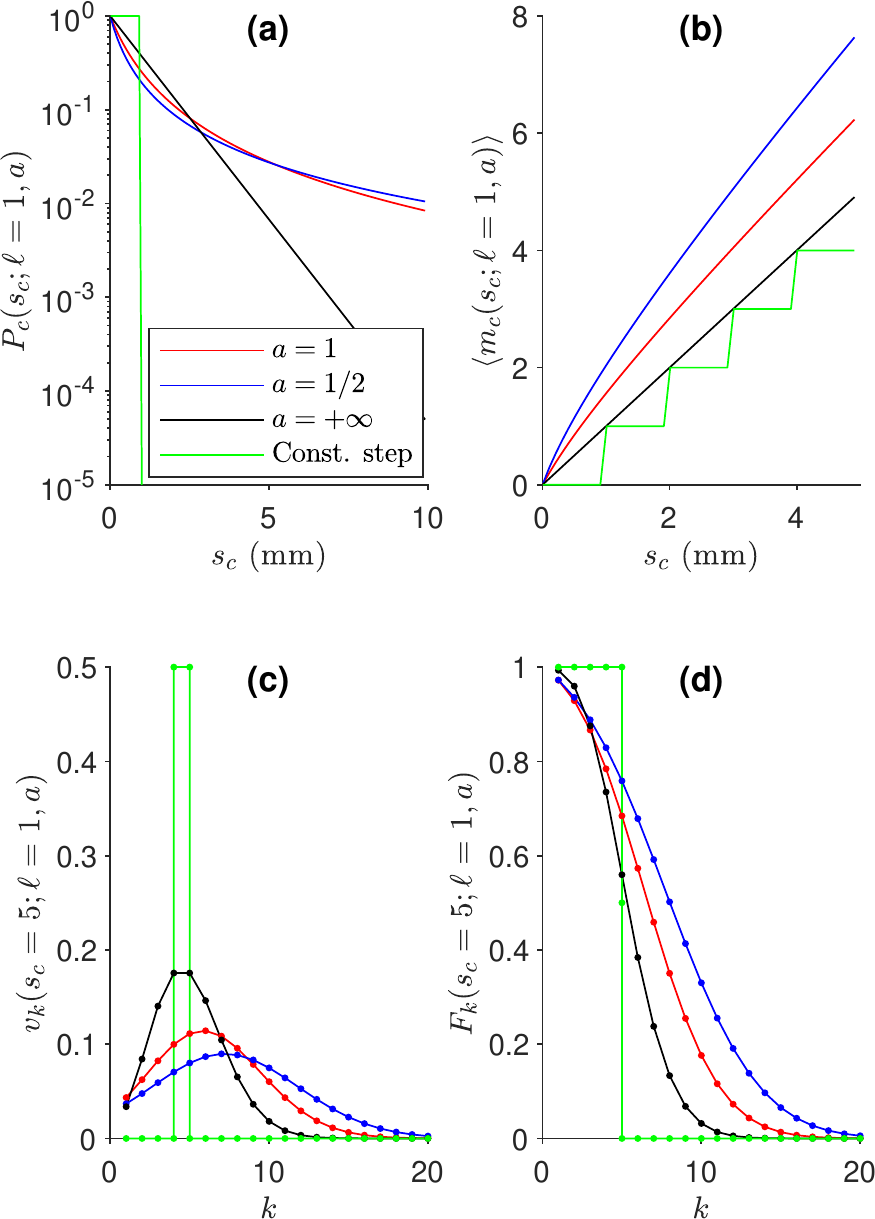}}
\caption{Probabilistic functions 
(see Sec.~\ref{sec:Basic MC tools in ART}.\ref{sec:Extracting the physical properties from the PDF}) 
characterizing the models chosen as a tutorial cases 
(see Secs. \ref{sec:Anomalous transport: an explanatory example}.\ref{sec:a=1}, 
\ref{sec:Anomalous transport: an explanatory example}.\ref{sec:a=1/2},
\ref{sec:Anomalous transport: an explanatory example}.\ref{sec:a=Inf}) and 
\ref{sec:Anomalous transport: an explanatory example}.\ref{sec:ConstantSteps}).
}
\label{fig:FigRenewal}
\end{figure}

In Fig.~\ref{fig:FigRenewal} 
the power law [Eqs.~(\ref{eq:PPPP1}) and (\ref{eq:PPPP2}) ], 
for $a=1$ and $a=1/2$, and the constant step [Eq.~(\ref{eq:PPPP3})] models, are compared with the special case $a=+\infty$ [Eq. (\ref{eq:PPPP4})],
related to the classical Beer–Lambert–Bouguer's law.
For simplicity we have chosen the case $\ell=1$, but the general considerations remain valid
for other $\ell$ values. 

In Fig. \ref{fig:FigRenewal}a the classical model for $P_c(.)$ appears as a straight line 
(logarithmic scale). 
It is known from renewal theory that this model is representative of a Markovian or memoryless process.
This means that a future photon step length only depends on the present state and not
on the past history of the photon propagation.
The remaining models appearing in  Fig. \ref{fig:FigRenewal}a,
that deviate from this behavior, are not memoryless. 
This explains in part why it may be more complex to express these models through an (integro)-differential equation formalism.
In general, $P_c(.)$ may be assessed experimentally by
the measurement of the so called ballistic photons, while the remaining functions 
in Fig. \ref{fig:FigRenewal} may only be indirectly derived by means of a mathematical
procedure.
Figure \ref{fig:FigRenewal}a clearly shows that the power low curves for ART go slower to zero compared to the classical case.
This means that there is a higher probability for a photon to 
suddenly undergo very long steps. 
This may be e.g. the case of photons that pass through a non-homogeneous distribution of clouds. 

Figure \ref{fig:FigRenewal}b represents the average number of scattering events (or steps), $\langle m_c(.) \rangle$, along a path of length $s_c$.
The figure highlights the fact that even if in general the average step is always $\ell$ for the power law  [Eq. (\ref{eq:ExpModelc})] and the constant step [Eq. (\ref{eq:step})] models,
for the specific ART condition it is not possible to simply divide the path length $s_c$ by $\ell$ to find the $\langle m_c(.) \rangle$ value, as it can be done in the classical case [Eq. (\ref{eq:mcaInf})] (black color; straight line).
In fact, Fig. \ref{fig:FigRenewal}b shows that ART curves are not linear and thus a specific calculation must be performed [Eq. (\ref{eq:m})]. 
This particular behavior is generated by the ``memory'' of the ART models.
Function $\langle m_c(.) \rangle$ is extremely important because renewal theory tells us that it uniquely determines the process generating the photons' steps.

Figure \ref{fig:FigRenewal}c reports the probability $v_k(.,.)$ that k scattering events occur on a path of length $s_c=5$ mm. 
As expected, in the classical case $v_k(.,.)$ is a Poisson distribution [Eq. (\ref{eq:vkaInf})], due to the 
exponentially distributed photons steps.
The function $v_k(.,.)$ can be utilized to reproduce the statistical spatial distribution of the scattering sites inside the medium, and gives us an idea on how the medium is structured.

Figure \ref{fig:FigRenewal}d reports the probability $F_k(.)$ that the sum of the first $k$ 
photon steps does not exceed $s_c=5$ mm.
It can be seen that the classical case has smaller values than the ART power law (apart for the cases $k=1, 2$).
The exceptions for $k=1, 2$ can be intuitively explained by looking e.g. 
Fig. \ref{fig:FigRenewal}a 
from which we can deduce e.g. that the probability to have a long step larger than 5 mm is higher for the classical case. 

Thus, Fig. \ref{fig:FigRenewal} gives us the main features of a given ART model compared to the classical one. 
The above (semi)-analytical approach may be sometimes 
useful for the interpretation of the MC simulations. 

\subsection{MC results} 

The results of the MC simulations, for the three cases and the geometrical model appearing in
Fig. \ref{fig:figureTwoSperes}, are summarized in Table~\ref{tab:shape-functions}. 
\begin{table*}[htbp]
\centering
\caption{Explanatory tests of the IP for the ART theory. 
Parameters of the simulations: $r_{out}=5$ mm, $r_{in}=2$ mm, $\ell_{out}=1$ mm, 
$\ell_{in}=3$ mm, $\langle s \rangle_\textrm{theory}=6.\bar6$ mm (for the remaining parameters see text).
PL: power law; CS: constant step; BLB: Beer–Lambert–Bouguer's.
Values in parenthesis are the standard error of the mean.}
\begin{tabular}{ccccc}
\hline
pdf model & $\langle s \rangle^\textrm{Fwd}_\textrm{MC}$ & $\langle s \rangle^\textrm{Fwd}_{\textrm{MC}_\textrm{no-reciprocity}}$ 
    & $\langle s \rangle^\textrm{Adj}_\textrm{MC}$ & $\langle s \rangle^\textrm{Adj}_{\textrm{MC}_\textrm{no-reciprocity}}$\\ 
\hline
PL $(a=1)$    & $6.6666 \; (0.0004)$ & $5.4777 \; (0.0002)$ & $6.6668 \; (0.0002)$ & $5.4242 \; (0.0002)$\\
PL $(a=1/2)$  & $6.6667 \; (0.0002)$ & $4.9731 \; (0.0002)$ & $6.6667 \; (0.0002)$ & $4.9030 \ (0.0002)$\\
CS   & $6.6666 \; (0.0004)$ & $9.3866 \; (0.0004)$ & $6.6668 \; (0.0004)$ & $9.6856 \ (0.0004)$\\   
BLB $(a=+\infty)$ & $6.6668 \; (0.0003)$ & $6.6664 \; (0.0002)$ & $6.6666 \; (0.0003)$ & $6.6668 \; (0.0003)$\\
\hline
\end{tabular}
\label{tab:shape-functions}
\end{table*}
The theoretical mean  pathlength that we must reproduce with the MC simulations is 
$\langle s \rangle_\textrm{theory}=\frac{4}{3} r_{out}$ mm [Eq. (\ref{eq:IP})].

From Table~\ref{tab:shape-functions} it clearly appears that, $\langle s \rangle^\textrm{Fwd}_\textrm{MC}$
perfectly reproduces the theoretical $\langle s \rangle_\textrm{theory}$ IP-value (within the statistical error) for all cases, i.e., the power law, the constant step and the classical model.
Moreover, $\langle s \rangle^\textrm{Fwd}_\textrm{MC}=\langle s \rangle^\textrm{Adj}_\textrm{MC}$ 
(within the statistical error).

However, if  the condition for the RL is not
satisfied, i.e., if we neglect the special behavior for 
uncorrelated scattering events and set $p_u(.)$ equal to $p_s(.)$, then the MC simulations generates results in disagreement with the laws of optics,
i.e., $\langle s \rangle^\textrm{Fwd}_\textrm{MC$_\textrm{no-reciprocity}$}  \neq \langle s \rangle^\textrm{Adj}_\textrm{MC$_\textrm{no-reciprocity}$}$,
$\langle s \rangle^\textrm{Fwd}_\textrm{MC$_\textrm{no-reciprocity}$}\neq \langle s \rangle_\textrm{theory}$ and 
$\langle s \rangle^\textrm{Fwd}_\textrm{MC$_\textrm{no-reciprocity}$}\neq \langle s \rangle_\textrm{theory}$,
 when $a\in\{1,1/2\}$.
The only exception being the classical case ($a=+\infty$), where all results remain valid (last line of Table~\ref{tab:shape-functions}).
This is obviously due to the fact that in this case $p_u(.)=p_s(.)$ [Eq. (\ref{eq:classipdfsu})].

Considering the geometry of the problem, the inequality 
$\langle s \rangle^\textrm{Fwd}_\textrm{MC$_\textrm{no-reciprocity}$}  \neq \langle s 
\rangle^\textrm{Adj}_\textrm{MC$_\textrm{no-reciprocity}$}$, for the ART,
immediately tells us that the light reaching the detector in the forward direction
do not represent an isotropic and uniformly distributed radiance. 
In other words, one of the conditions for the IP is not satisfied for the adjoint
case. 
As expected, in the classical case this problem does not subsist, i.e., 
$\langle s \rangle^\textrm{Fwd}_\textrm{MC$_\textrm{no-reciprocity}$}  = \langle s 
\rangle^\textrm{Adj}_\textrm{MC$_\textrm{no-reciprocity}$}$.

In Table~\ref{tab:shape-functions}, the  data are exact up to the third decimal but,
obviously, if we increase the number of photons trajectories per simulation, we can
also increase the precision of the results at any desired level.


\section{Discussion and conclusions}
\label{sec:Discussion and conclusions}

In Sec. \ref{sec:Anomalous transport: an explanatory example} and \ref{sec:Results} 
we have seen, through some tutorial example, the importance to distinguish between correlated and uncorrelated scattering events,
through the pdfs $p_c(.)$ and $p_u(.)$.
In fact, this approach not only satisfies (by construction of the theory) 
the RL for photon propagation, 
but it is also mandatory if one want to satisfy the universal IP. 
 
In Table~\ref{tab:shape-functions} we have also seen that 
there is an exception for the case  
$a=+\infty$, corresponding to the classical case where photons propagate
under the Beer–Lambert–Bouguer's law. 
In fact, in this special case (the only one) $p_u(.)=p_s(.)$, 
and as a consequence the results remain exact in all cases 
(Table~\ref{tab:shape-functions}; last line).

In practice, the ART theory is a generalization of the classical one,
the latter appearing as a particular case.
Thus, the ART MC method can be applied to solve classical problems in photon transport,
e.g., in biomedical optics.
This may greatly simplify the writing of MC codes (the photon step just stops at the boundary, 
before the next step), because it does not necessitate to treat 
the photons reaching the boundaries with more complex algorithms as it is classically done
\cite{ref:Wangetal95}.

\begin{figure}[htbp]
\centering
{\includegraphics[width=\linewidth]{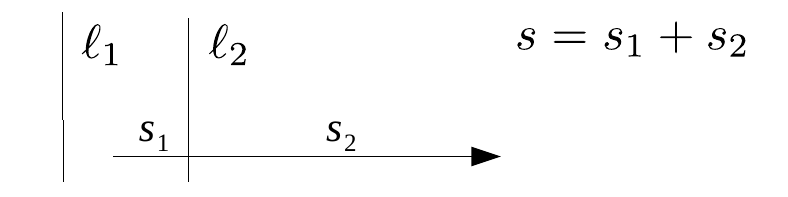}}
\caption{Schematic of a photon that crosses a boundary separating
two regions with different mean free path.}
\label{fig:figureTwoSperes2}
\end{figure}

But then, do we have two different algorithms to treat the photons interactions with the boundaries  in classical MC ? 
Which is the right one?
Actually,  the classical and the ART based MC methods give the same results.
Intuitively, it is probably hard to see that the two approaches are equivalent.
For this reason, another simplified example might maybe help the reader to have a more concrete idea.
Let be a medium with only one internal boundary and mean free paths $\ell_1$ and $\ell_2$.
Let generate a photon step,
$s=s_1+s_2$, as in Fig. \ref{fig:figureTwoSperes2}.
It is well known that in classical MC the random step $s$ is obtained as
\begin{equation}
s=\begin{cases}
-\ell_1\ln\xi; & \mbox{if } s\le s_1 \\
\frac{\ell_2}{\ell_1}(-\ell_1\ln\xi-s_1)+s_1; & \mbox{if } s \mbox{ computed above} > s_1,
\end{cases}
\label{eq:sClassRand}
\end{equation}
where $\xi\in(0,1)$ is a uniform distributed random variable.

In the same way, as it was explained in this tutorial, in ART the MC random step $s$ is generated as
\begin{equation}
s=\begin{cases}
-\ell_1\ln\xi_1; & \mbox{if } s\le s_1 \\
-\ell_2\ln\xi_2+s_1; & \mbox{if } s \mbox{ computed above} > s_1,
\end{cases}
\label{eq:sARTRand}
\end{equation}
where $\xi_1,\xi_2\in(0,1)$ are independent uniform distributed random variables.
It easy to see (e.g. numerically) that Eqs. (\ref{eq:sClassRand}) and (\ref{eq:sARTRand}) generate 
random $s$ with the same pdf.
This equivalence is valid in general for any choice of complex geometries, 
number of boundaries, optical properties, etc.
This is the reason why, in the case of classical simulations where Eq. (\ref{eq:classicalpdf}) holds,  ART and the classical  MC approaches are equivalent.

At this point, without entering to much in too technical details, it is maybe worth making a general consideration concerning the RL and the IP.
In fact, historically, the pdf $p_u(.)$ has been derived independently for the RL and for the IP theory.
Each theory having its own independent (but compatible) physical assumptions allows us to assess $p_u(.)$.
In this didactical contribution, we have chosen a point of view that
is not historical but that should better highlight the relationship existing between RL and IP.
In fact, the RL is a more ``general'' law than the IP. This, because RL holds e.g. also for
any light source, detector configuration, optical properties values, or phase function choice.
Thus, the pdf for a step starting from a fixed position must be $p_u(.)$, if we want RL
to be satisfied.

On the other hand, historically, the pdf $p_u(.)$ necessary to satisfy the IP
  has also been derived independently \cite{ref:Mazzolo2014} 
(the obtained $p_u(.)$ equals the one satisfying the RL),
but by using different hypothesis (actually, a subset of conditions of the RL).
We realize here that this derivation for the IP is not strictly necessary, because
$p_u(.)$ is already given by the RL.
In practice, it is possible to demonstrate the IP by considering $p_u(.)$ already known.
We invite the interested reader to revisit the original papers on IP
(e.g. Ref. \cite{ref:Mazzolo2014})
with this point of view.
This might help to have a more coherent vision on the strong relationship
existing between RL and IP.

The few examples 
presented in this tutorial
had the aim to show how to generate MC simulations for any
$p_c(.)$ describing a desired medium.
However, we must be aware of the fact that to describe the exact physics 
of the investigated medium, the first moment of $p_c(.)$ must be finite [Eq. (\ref{eq:pupdf})].
From a practical point of view this constraint on $p_c(.)$ is not a real problem, 
because when simulating media representing actual physical systems, it is 
unlikely that photons goes to infinity without any interaction.
This keeps the average step length finite.

In conclusion, it may be worth to note that an MC code 
compatible with the ART theory may also be of a more general interest, than a simple technicality, for tissue optics.
In fact, it has been shown that some biological tissues, such as bone and lung, 
have a particular fractal-like structure \cite{ref:Madzin2008,ref:Czyz2017,ref:Lennon2015}.
As  reported in Ref. \cite{ref:Davis2011} by Davis and Mineev-Weinstein, (multi-)fractal structures may
produce anomalous photon propagation. 
This observation may open a new domain of exploration in biomedical optics,
and adapted MC codes may represent one of the important tools to address this 
interesting topic.

%
%


\end{document}